\documentclass[lnbip]{svmultln}
\def\basiceval#1{\the\numexpr#1\relax}
\usepackage{float} 
\usepackage{booktabs} 
\usepackage[utf8]{inputenc} 
\usepackage{amsfonts}
\usepackage{graphicx}
\usepackage{algorithm}
\usepackage{algpseudocode}
\usepackage{amsmath}
\usepackage[inline]{enumitem}
\usepackage{titlesec}

\usepackage{environ}
\usepackage{xstring}

\titlespacing*{\section}{0pt}{1\baselineskip}{1\baselineskip}

\usepackage{tabto}

\usepackage{makeidx}  
\begin{document}
\mainmatter              
\title{Personalisation of d'Hondt's algorithm and its use in recommender ecosystems}
\titlerunning{Personalisation of d'Hondt's algorithm}  
%
\author{Stepan Balcar \and Ladislav Peska \and Peter Vojtas}
\authorrunning{Stepan Balcar et al.}   
%
\tocauthor{Stepan Balcar, Ladislav Peska, Peter Vojtas}
\institute{Charles University, Faculty of Mathematics and Physics, Prague, Czech Republic,\\
\email{stepan.balcar |  ladislav.peska | peter.vojtas @matfyz.cuni.cz}}

\maketitle              

\begin{abstract}        
In the area of recommender systems, we are dealing with aggregations and potential of personalisation in ecosystems. Personalisation is based on separate aggregation models for each user. This approach  reveals differences in user preferences, especially when they are in strict disagreement with global preferences. Hybrid models are based on combination of global and personalised model of weights for d'Hondt's voting algorithm. This paper shows that personalisation combined with hybridisation on case-by-case basis outperforms non-personalised d'Hondt's algorithm on datasets RetailRocket and SLANTour. By taking into account voices of minorities we achieved better click through rate.


\end{abstract}
In recommender systems community, heterogenity is a well known concept. Robustness of ensembles of base recommenders offers a general tool which increases diversity, ensures fairness and allows processing of multimodal data. In the area of online aggregators based on the principle of rewards,  as an alternative to the Multi Armed Bandits are considered voting algorithms e.g. d'Hondt's \cite{balinski2010fair}, D21-Janecek \cite{book} which are often used as proportional aggregators.

As recently summarized by Kangas et al. \cite{kangas2021recommender} in the context of Booking.com, personalisation should be done in clusters e.g. Multi Armed Bandits. Our approach seeks to take advantage of the collaboration between the global and fully personalised model. We do not use context-based clustering because clusters of interest do not necessarily have to correspond to clusters of ideal weight aggregation models. Instead, we are giving fully-personalised models enough time to learn or take information from more experienced models. The benefits of the hierarchy of aggregators in ecosystems have already been proven \cite{iccai2022Balcar}. As a result, this hierarchical approach to personalisation doesn't hurt but instead improves the performance of the recommending ecosystem.  In this paper it's shown that surrender of short timeslice of the dataset will prevent ignoring of minority voices.

\subsubsection{Main contribution:} Full personalisation significantly increases relevance. Hierarchical heterogeneous portfolios are generalised on the level of voting-based aggregational data models. Potential synergy of global and personalised models is investigated as well.

\section{Personalisation of d'Hondt's algorithm}

As a way of performing aggregation, d'Hondt's algorithm has proved its merits \cite{balcar2022rank}. However, in the cited paper personalisation was performed only during integration of implicit negative feedback (INF). There was only one model of aggregation for all the users. In this paper, we are investigating full or hybrid personalisation of d'Hondt's algorithm on the level of weight model. Full personalisation means that each user has their own aggregation model. We assumed that full personalisation will need special pre-training of aggregators and will also face many challenges when adding new users due to cold-start problem.

\begin{figure}[ht]
    \centering
    \includegraphics[width=\columnwidth]{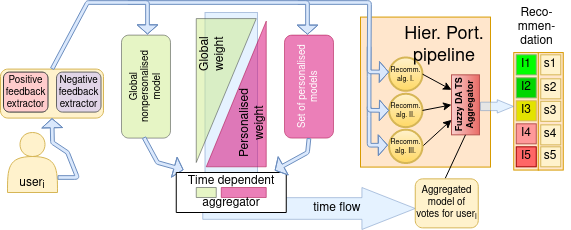}
    \caption{Hybrid aggregation model. It aggregates personalised and global model of votes based on threshold changing in time.}
    \label{fig:hierarchicalPortfolioDataModel}
\end{figure}

In an effort to cope with these issues, we designed various approaches, mostly based on hybrid portfolios. They combine global and personalised model, where weight ratio of global and personalised model is changing with time. At the beginning, global model is dominating (see Fig. \ref{fig:hierarchicalPortfolioDataModel}). Consecutively its importance is overtaken by personalised model, which is unique for each user. In one variant of the solution of cold start problem, personalised model for a new user was initialized by a duplicate of the model of the user with the highest click count. We experimented further with normalization of responsibility for recommended items.

\section{Experiments design}

\begin{figure}[ht]
    \centering
    \includegraphics[scale=0.28]{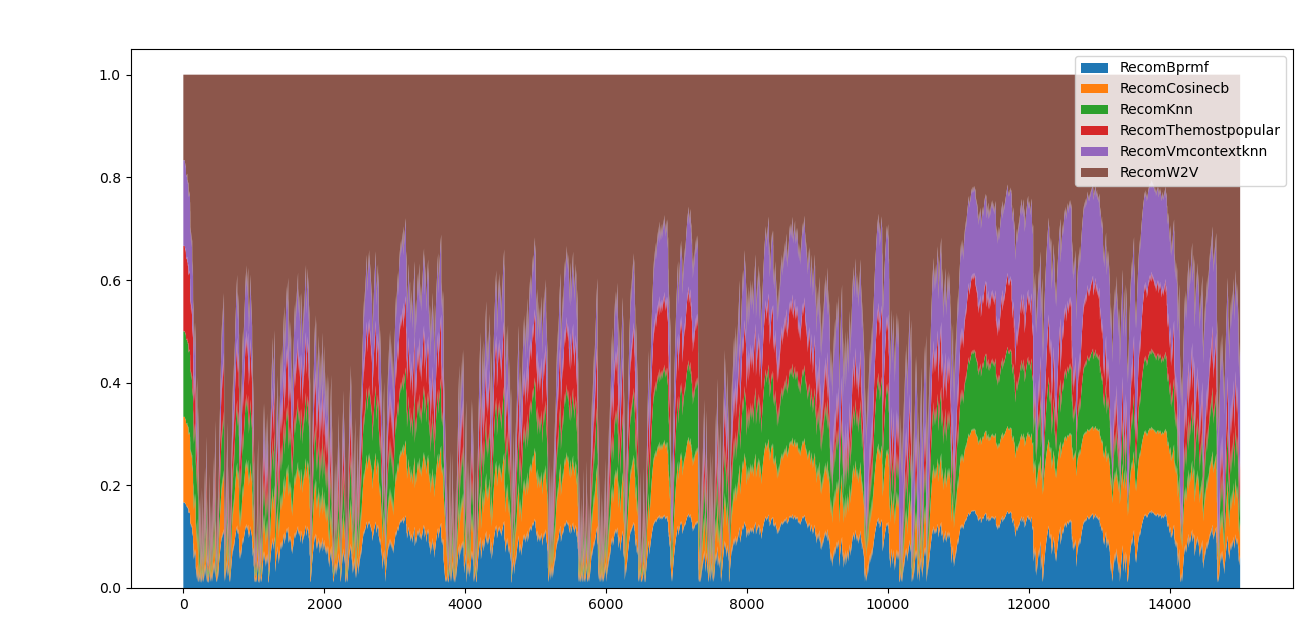}
    \caption{Weight development of model of votes.}
    \label{fig:weightDevelopment}
\end{figure}

The basic intention was an attempt to fully personalise d'Hondt's model in simulated online environment. In other words, we are training base recommenders, but not aggregators. Weights inside aggregators are initialized uniformly so that each method obtains the same number of initial votes. In each step, modification of weights of aggregators (see Fig. \ref{fig:weightDevelopment}) is based on reward-and-punishment principle. If an item receives a click, the method, which has the most responsibility for recommending it, receives a reward. On the other hand, if user doesn't click on the recommended item, weights of methods responsible for its recommendation are penalized. In this manner, methods in an online environment get greater or lesser opportunity to make contributions to the final aggregated list.

Our hybridisation of D'Hondt's models of votes is based on idea of multiplicative weights update \cite{v008a006}. We attempted to combine information stored in the global model with user-specific knowledge coming from developing personalised model. This combination was based on an assumption that learning of personalised model for each user will worsen the performance (some interactions in sequential evaluation will need to be sacrificed to the training of personalised model) and training will take longer. Experiments consist of parallel development of both global and personalised model (see Fig. \ref{fig:hierarchicalPortfolioDataModel}). We designed other variants of model union as well. In one variant we ignore personalised model for users who have less than three clicks.

For evaluation of designed models, we use sequential evaluation \cite{idomaar}. It walks through the dataset and at each event, it recommends items to the current user. Recommendation is counted as a (simulated) click if it's both relevant (contained in item window) and was noticed (noticeability is described by user behaviour model). Each dataset was sorted by timestamp and its first half was used for training of base recommenders. The last 10\% was used for sequential evaluation. Window size was set to 5 items. Base recommenders recommend 100 items and aggregator returns a list of 20 items. Users' ability to observe item \cite{balcar2022rank} on $kth$ position is simulated in two ways: At first, static $P_{stat08}(noticed|k) = 0.8$ or $P_{stat06}(noticed|k) = 0.6$ (probability of noticing is equal to 0.8 or 0.6 for all the items) and at second, linear $P_{lin0901}$ (probability that user has noticed the item decreases from 0.9 to 0.1 linearly with item position in the list).

Experiments were performed on datasets RetailRocket\footnote{https://www.kaggle.com/datasets/retailrocket/ecommerce-dataset} and medium-sized Czech travel agency SLANTour\footnote{https://www.slantour.cz/} - 12 months of the user visit events (mid January till late December 2020). We chose them because they both represent logs of online experiments and contain repeated user-item interactions. RetailRocket contains 2.7M and SLANTour 200K events. To make computations less time consuming, we filtered out of RetailRocket only users who had more than 50 interactions with the system. For SLANTour, we kept all the events. Each of these datasets has a different structure. SLANtour contains both more users with just one interaction and more users with thousands of interactions, whereas in RetailRocket, distribution of event counts is close to uniform.

In each experiment, we used the same base recommenders: the most popular recommender, word2vec \cite{NIPS2013_9aa42b31} recommender, content-based cosine similarity recommender \cite{10.1145/3372923.3404781}, BPRMF (matrix factorization-based), item-based KNN recommender \cite{10.1007/s11257-017-9194-1} and session KNN recommender \cite{10.1007/s11257-017-9194-1}.

We first aimed at using online development of weights of base RS for selection of optimal configuration of weights. Fig. \ref{fig:weightDevelopment} shows weight development during model training. It's clear that weights are oscillating, which led us to the conjecture that "trends" for various users are interfering with each other. That was our main motivation to personalise whole portfolio model for each user.

\section{Results of experiments}

In previous section we have described variants of full or hybrid personalisation. We have considered all these variants as candidates for hyperparameter tuning. In general, variants which skipped underdeveloped personalised models performed better for linear behaviour model and worse for static ones. Normalization of responsibility always improved the results, with a single exception of static behaviour model for SLANTour.
From the table of the best results (see Table \ref{tab:results}) it follows that for RetailRocket, hybrid models outperform both non-personalised and fully personalised models. However, the case is different for SLANtour dataset - the personalised models win at all counts. This contradicts most common assumptions about personalisation. We supposed that bigger dataset will have a greater chance to profit from online-learnt personalised model and that results of full personalisation will be worse for SLANTour, because simulation takes less time there. It turns out that results are only slightly influenced by the size of the dataset, over which we are performing sequential evaluation. What matters most are user behaviour model, click frequency and learning rate for weights in d'Hondt's model.

\begin{table}[h]
\centering
\begin{tabular}{l||rrrrrr}
 RS / Evaluation variant & \rotatebox{48}{RR; lin0901} & \rotatebox{48}{RR; stat06} & \rotatebox{48}{RR; stat08} & \rotatebox{48}{ST; lin0901} & \rotatebox{48}{ST; stat06} & \rotatebox{48}{ST; stat08} \\
\hline \hline
FDHondt Non-personalised     & 0.1149           & 0.1151             & 0.1317             & 0.1642             & 0.1574          & 0.1796 \\
\hline
Full-personalized            & 0.1123           & 0.1138             & 0.1145             & \textbf{0.1656}    & \textbf{0.1672} & \textbf{0.1956} \\
Hybrid-personalized          & \textbf{0.1188}  & \textbf{0.1180}    & \textbf{0.1332}    & 0.1482             & 0.1504          & 0.1849 \\ 
\hline
\end{tabular}
\caption{\label{tab:results} Evaluation results. Columns correspond to the evaluation variants and values to the click-through counts. "Lin" and "stat" denotes linear and static noticeability respectively.}
\end{table}

After analyzing results of hybrid models (which give us an opportunity to compare global and personalised models evolving in parallel), it shows that online weights development is not so time-consuming and doesn't harm the results significantly (even in the case of full personalisation). The main factor contributing to results is whether weight distribution of global model agrees with weight distribution of the majority of personalised models.

In SLANtour experiment, when it comes to the global model, SessionKNN dominates over BPRMF. During sequential evaluation, both of these methods are trying to come to the top, but already before 1000th iteration SessionKNN prevails over BPRMF. In the beginning, situation is the same for personalised models. But after 4000 iterations most of the personalised models break the mould and take their own way, which makes BPRMF rise to the top.

In RetailRocket experiment we would expect fully personalised models to outperform the global ones. Simulation takes more time and offers more opportunities to utilize model which is fine-tuned for the specific user. Analysis of results has shown that personalised models have the same weight distribution as global models most of the time. Since they agree on the most significant base recommender, their aggregation will boost the click through rate \cite{russel2010}.

\begin{table}[h]
\centering
\begin{tabular}{l||l|l}
 Model Type / Dataset & RR & ST  \\
\hline
Global         & cosineCB           & session KNN             \\
\hline
Personalised   & cosineCB           & BPRMF                   \\
\hline
\end{tabular}
\caption{\label{tab:mostSignificant}Most significant base RS}
\end{table}

\section{Conclusion and future work}

Extension of personalisation of recommendation ecosystem to the level of data models - in particular, aggregators of d'Hondt's proportional voting algorithm - proves itself to be effective in increasing relevance of recommendations. Influence of aggregations doesn't just solve the problem of selection of base RS for specific domain. It also can be seen that weight distributions, which differ significantly across userbase, bring improved relevance.

Personalised models are penalized because of time-consuming nature of separate training of each user model, but nevertheless achieve the best values for SLANTour dataset. For the RetailRocket dataset, hybrid d'Hondt's data model achieved the first place because it has reached synergy between global and personalised model.

This paper shows that full personalisation of d’Hondt’s algorithm on the level of weight model not only increases relevance of recommendations, but gives voice to preferences of minorities.  On the other hand, hybrid personalisation allows to make use of synergy between global and personalised models. 

\subsubsection{Future work:}

In our future work, we would like to research the reason for oscillations in the global model (see Fig. \ref{fig:weightDevelopment}) and causes of discrepancies between global and personalised models. For performance improvement, adaptive clustering could help. The main scientific potential of aggregations lies in possibility of revealing consequences of ignoring voices of marginalised minorities. We intend to extend our work to more data domains and find better ways of coping with cold-start problem as well.

\subsubsection{Acknowledgement:}

The work on this paper has been supported by Czech Science Foundation project  GACR-19-22071Y  and  by  Charles  University  grants SVV-260588 and  GAUK-1026120. Source codes and trained models are available at \url{https://github.com/sbalcar/HeterRecomPortfolio}.

%
%
%

\begin{thebibliography}{10}

\bibitem{v008a006}
Sanjeev Arora, Elad Hazan, and Satyen Kale.
\newblock The multiplicative weights update method: a meta-algorithm and
  applications.
\newblock {\em Theory of Computing}, 8(6):121--164, 2012.

\bibitem{balcar2022rank}
Stepan Balcar, Vit Skrhak, and Ladislav Peska.
\newblock Rank-sensitive proportional aggregations in dynamic recommendation
  scenarios.
\newblock {\em User Modeling and User-Adapted Interaction}, pages 1--62, 2022.

\bibitem{iccai2022Balcar}
Štěpán Balcar, Ladislav Peška, and Peter Vojtáš.
\newblock Hierarchical portfolios in recommender ecosystems.
\newblock {\em ICCAI 2022, Tianjin, China (not published yet)}, 2022.

\bibitem{balinski2010fair}
Michel~L Balinski and H~Peyton Young.
\newblock {\em Fair representation: meeting the ideal of one man, one vote}.
\newblock Brookings Institution Press, 2010.

\bibitem{NIPS2013_9aa42b31}
Tomas~Mikolov et~al.
\newblock Distributed representations of words and phrases and their
  compositionality.
\newblock In {\em Advances in Neural Information Processing Systems},
  volume~26. Curran Associates, Inc., 2013.

\bibitem{10.1007/s11257-017-9194-1}
Dietmar Jannach, Malte Ludewig, and Lukas Lerche.
\newblock Session-based item recommendation in e-commerce: On short-term
  intents, reminders, trends and discounts.
\newblock {\em User Modeling and User-Adapted Interaction},
  27(3–5):351–392, dec 2017.

\bibitem{kangas2021recommender}
Ioannis Kangas, Maud Schwoerer, and Lucas~J Bernardi.
\newblock Recommender systems for personalized user experience: Lessons learned
  at booking. com.
\newblock In {\em Fifteenth ACM Conference on Recommender Systems}, pages
  583--586, 2021.

\bibitem{10.1145/3372923.3404781}
Ladislav Peska and Peter Vojtas.
\newblock {\em Off-Line vs. On-Line Evaluation of Recommender Systems in Small
  E-Commerce}, page 291–300.
\newblock Association for Computing Machinery, New York, NY, USA, 2020.

\bibitem{book}
Marcin Pomarański, Dorota Maj, Maria Marczewska-Rytko, and Kamil Aksiuto.
\newblock {\em Civic Participation in the Visegrad Group Countries after 1989}.
\newblock 06 2018.

\bibitem{russel2010}
Stuart Russell and Peter Norvig.
\newblock {\em Artificial Intelligence: A Modern Approach}.
\newblock Prentice Hall, 3 edition, 2010.

\bibitem{idomaar}
Mario Scriminaci, Andreas Lommatzsch, Benjamin Kille, and Frank Hopfgartner~et
  al.
\newblock Idomaar: A framework for multi-dimensional benchmarking of
  recommender algorithms.
\newblock In {\em RecSys'16 Posters}, volume 1688 of {\em CEUR-WS}, 2016.

\end{thebibliography}
%

\end{document}